\documentclass{aa}
\usepackage{graphicx}
\def\deg{$^\circ$}

\def\***#1{***{\scshape #1}***}

\begin{document}

\sloppypar

   \title{First results from TOO observations of the Aql X-1 field
with INTEGRAL\thanks{Based on observations with INTEGRAL, an ESA project with instruments
    and science data centre funded by ESA member states (especially the PI
    countries: Denmark, France, Germany, Italy, Switzerland, Spain), Czech
    Republic and Poland, and with the participation of Russia and the USA.}}

   \author{S. Molkov, A. Lutovinov, S. Grebenev }

   \offprints{molkov@hea.iki.rssi.ru}

   \institute{Space Research Institute, Russian Academy of Sciences,
              Profsoyuznaya 84/32, 117810 Moscow, Russia
            }
  \date{}

  \authorrunning{Molkov et al.}
  \titlerunning{}
        
  \abstract{ We present results of observations of the Aql X-1 field
  performed in March-April 2003 with the INTEGRAL observatory. This
  TOO (Target Of Opportunity) INTEGRAL
  observations was initiated upon receiving an 
  indication from the ASM/RXTE that the source started an outburst. Thirteen
  X-ray sources were detected by the INTEGRAL imagers, JEM-X and IBIS, during
  these 
  observations.  We present a preliminary spectral and timing analysis for
  several bright sources in the field, Aql X-1, X1901+03, 4U1907+097,
  XTE J1908+094 and X1908+075. We also detected two X-ray bursts from 
  Aql X-1 near the end of the general outburst episode.
   }
\maketitle
%

\section{Introduction}

The hard X-ray emission has long been considered to be a unique property of
the Black Hole (BH) binaries. Later, however, it was found that X-ray
bursters can also be detected in the hard X-ray band (Barret \& Vedrenne
1994; Churazov et al. 1995; Pavlinsky et al. 2001). Moreover, some Neutron
Star (NS) binaries are often found in the spectral state with the enhanced
X-ray flux (e.g. Zhang et al 1996), which is similar to the so-called low
spectral state of Cyg~X-1 (Hasinger \& van der Klis 1989; Yoshida et al.
1993). The presence of the hard X-ray emission can be plausibly associated
with the low accretion rates. Therefore, one can expect transitions
  between the spectral states in X-ray transients at the beginning of
outbursts when the accretion rate increases and at the end of outburst when
it decreases until it falls below detectable level (Barret et al. 1996).

The Low Mass X-ray Binary (LMXB) Aql X-1 is ideally suitable for the study
of such spectral transitions. It is a well known soft X-ray transient
which demonstrates transitions from low to high flux states approximately
once a year (see the historical light curve shown in the bottom panel of
Fig.1).

The sky region around Aql X-1 contains several other bright X-ray sources
because it is projected on the Scutum arm of the
Galaxy. One of the most interesting sources in this field is the transient
X1901+03, which showed an outburst in January 2003 (Galloway et al. 2003a) for
only the second time since 1971 (Forman et al. 1976). During the 2003
outburst, X1901+03 showed coherent pulsations with a period of 2.763~s
(Galloway et al. 2003a), therefore this source is an accreting X-ray pulsar
in a binary system. Most of remaining sources are either pulsars or
black-hole candidates (4U1907+097, XTE J1908+094, X1908+075, SS433,
GRS1915+105, etc.) and are likely to be high-mass binaries (Grimm et
al. 2002).

\begin{figure}[t]
 \vspace{-0.5cm}
 \resizebox{90mm}{!}{\hspace{-0.8cm}\includegraphics{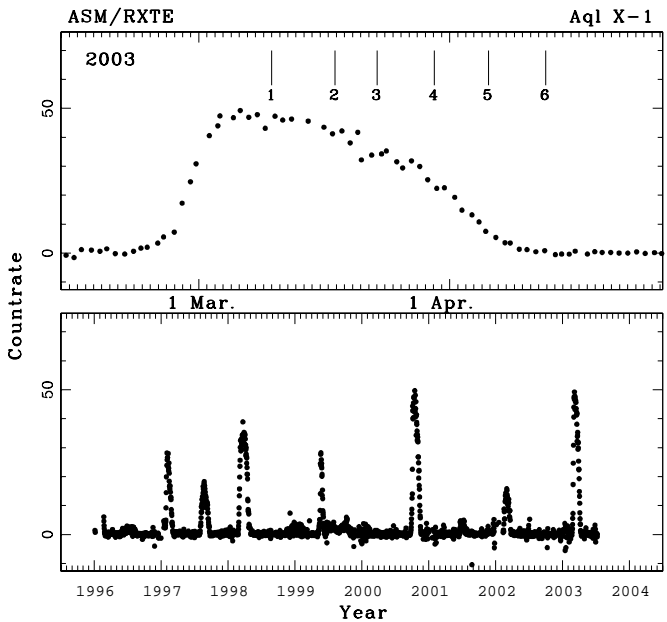}}
 \vspace{-5mm}
 \caption{Bottom: Historical light curve of Aql X-1 in the 2-12 keV energy
band, obtained by RXTE (data averaged over 1-day intervals). Top: Detailed
profile of the March, 2003 outburst. Vertical lines indicate moments of the
INTEGRAL observations.} 
 \label {aqlasm}
\end{figure}

In this paper, we present preliminary scientific results from the INTEGRAL
observations of the Aql X-1 field in March-April 2003.

\section{Observations and data reduction}

The present dataset was obtained in a campaign of TOO observations of Aql
X-1 during its outburst. Six approximately equally spaced (one per
five-six days, see Fig.1) observations were carried out from Mar 10 to Apr 13,
2003. Each observation consists of 25 pointings with a $\sim 2$ ksec
exposure, forming a $5\times5$ dithering pattern. The total integration time
during this campaign was approximately 300 ks. In this work, we present only
results from the Joint European X-ray Monitor (JEM-X, module 2; see Lund et
al. 2003) and the upper layer of the Imager on Board INTEGRAL (ISGRI/IBIS;
Ubertini et al. 2003). The detailed description of the instruments can be
found in these papers, but several main parameters like a field of view
(FOV), fully coded field of view (FCFOV), collecting area and energy range
are pointed here: $13.2$\deg in a diameter, $4.8$\deg in a diameter, 500
cm$^2$ and 3-35 keV for JEM-X, and 25\deg$\times$25\deg, 9\deg$\times$9\deg,
2600 cm$^2$ and 15-300 keV for ISGRI/IBIS, respectively.

Data reduction was performed using the IDAS 1.0 software distributed by the
INTEGRAL Science Data Center. This software package at present does not
allow one to carry out detailed spectral analysis of the ISGRI
data. Therefore, the spectral information for sources in the IBIS field of
view (FOV) was obtained from a comparison of their observed pulse height 
spectra with that of the Crab nebula as measured by INTEGRAL in February,
2003. The analysis of a set of the Crab observations
and preliminary analysis of hard X-ray emission from SAX J2103.5+4545
(Lutovinov et al. 2003) have shown that this method provides satisfactory
reconstruction of the source spectra and allows one to estimate the main
spectral parameters.

Our analysis of the Crab observations has shown a strong dependence of the
reconstructed source intensity on the off-axis angle. To reduce this
potential source of systematic errors, the spectral analysis of our sources
was performed using only the data from the fully coded part of FOV (FCFOV)
in the case if at least 20 suitable pointings were available for the source of
interest. We estimate the systematic
uncertainty of the flux for such data selection to be $\sim$10-20\% at all
energies. The instrumental 
teams independently estimate this uncertainty as at least $\sim$10\%. 
For the image reconstruction, we used the entire FOV of the IBIS telescope.

We note that at present, systematic uncertainties in the spectral
reconstruction dominate the statistical noise. Therefore, the usual reduced
$\chi^2$ is not a correct measure of the quality of the applied models. We
generally avoid quoting reduced $\chi^2$ for our spectral models.

Examination of the JEM-X data for Crab nebula shows that the current
calibration of the instrument response matrix is satisfactory in the 5--20
keV energy range. It was found, however, that while the spectral
shape is recovered well, the absolute source fluxes are systematically
underestimated. To correct this problem, we renormalized our JEM-X
spectra using the absolute fluxes in the 5--12~keV energy band provided by
the RXTE All-Sky Monitor simultaneously with the INTEGRAL observations.

\begin{figure}[t]
 \vspace{0.3cm}
 \resizebox{85mm}{!}{\hspace{-0cm}\includegraphics{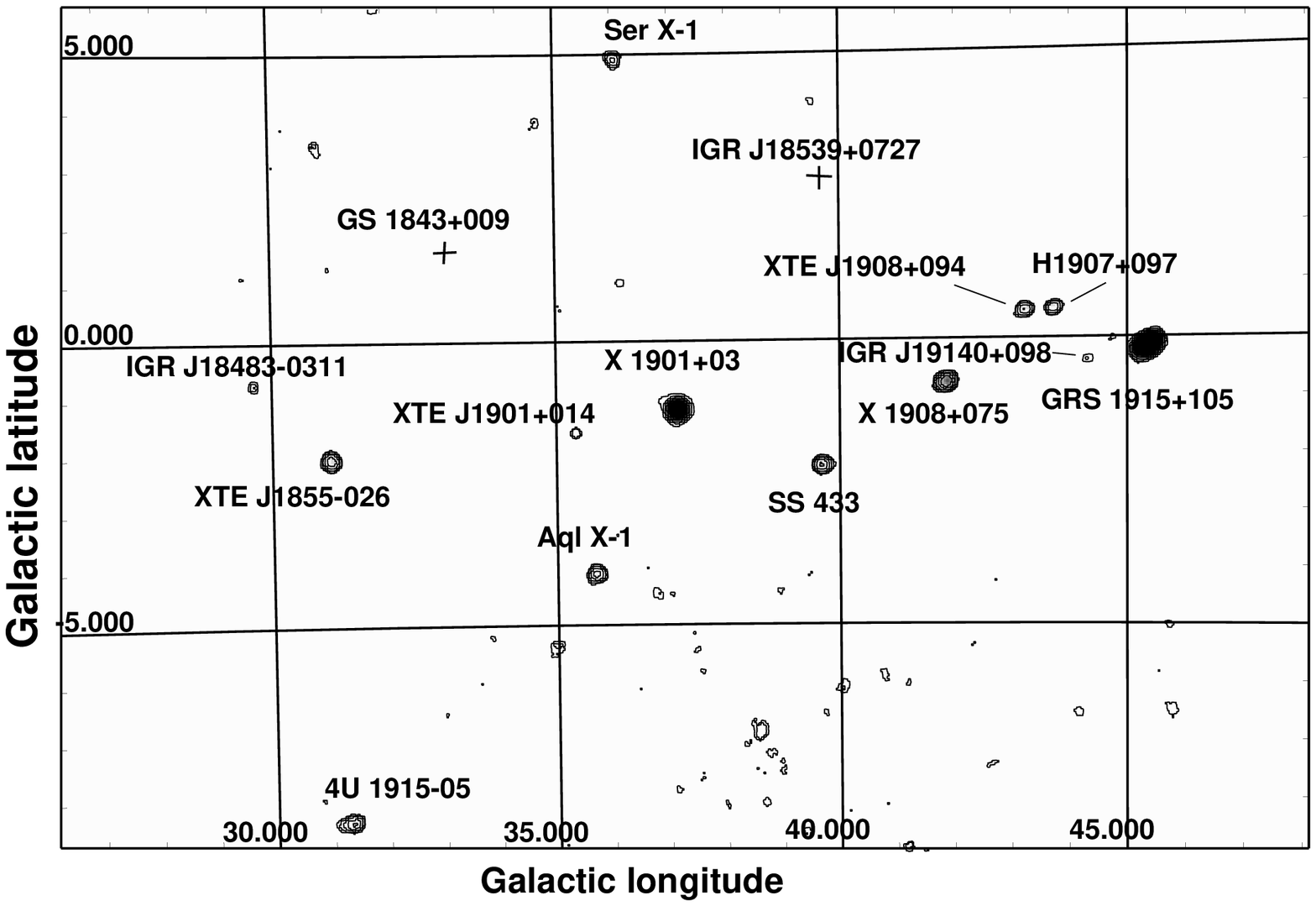}}
 
\vspace{-6mm}

 \resizebox{86mm}{!}{\hspace{-0.5cm}\includegraphics{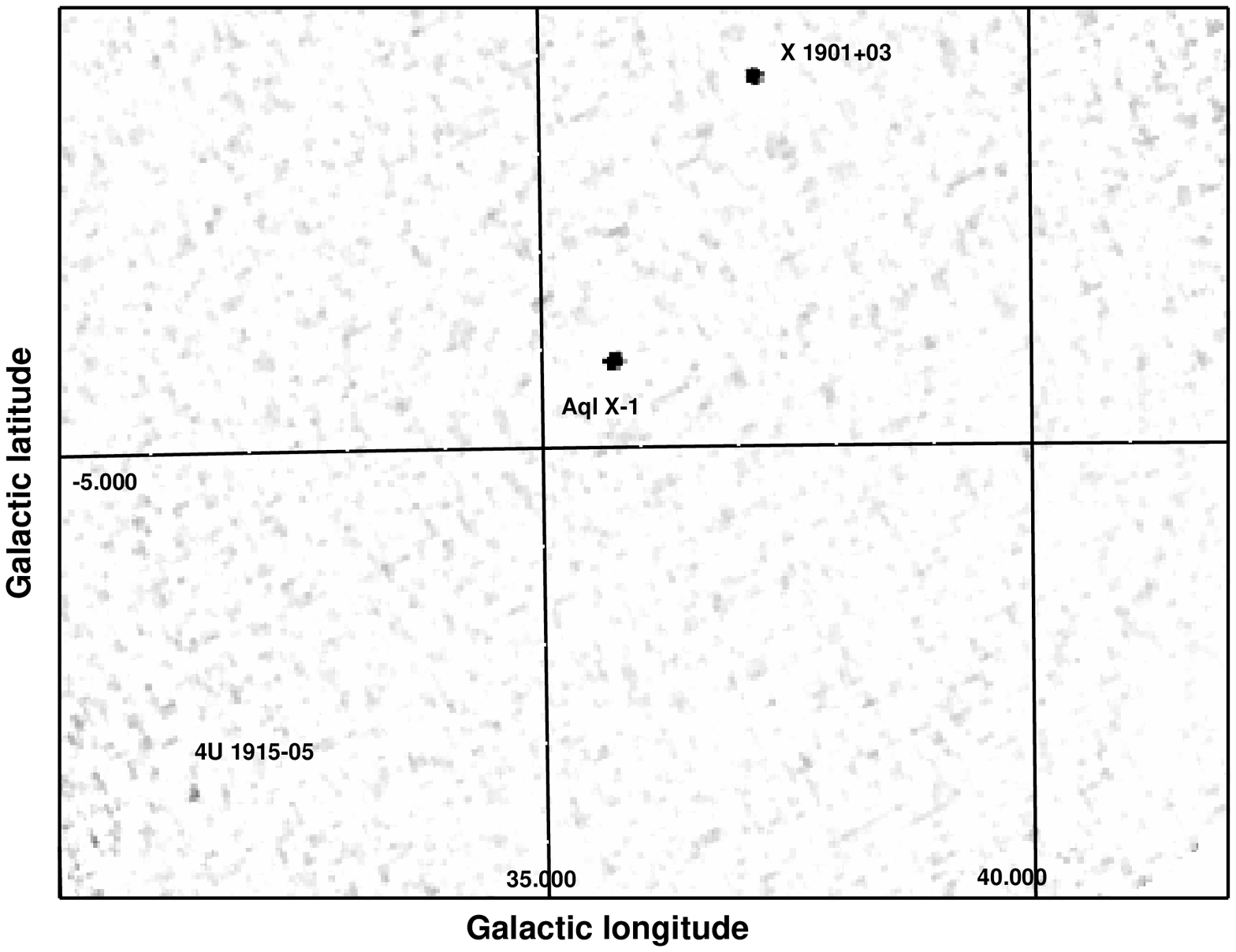}}
 \vspace{-5mm}
 \caption{X-ray images of the Aql X-1 sky region, obtained with
ISGRI/IBIS in the 25-50 keV energy band (upper panel) and JEM-X
in the 3-10 keV energy band (bottom panel) in March-April
2003. Contours on the ISGRI/IBIS map are given at the signal-to-noise levels of
2, 2.5, 3.3, 4.5, 6.2, ... 42 $\sigma$ (for the brightest sources
GRS1915+105 and X1901+03 contours begin from 4.5 $\sigma$). }
 \label {maps}
\end{figure}

\section{Results}

The JEM-X and ISGRI mosaic images reconstructed in the 5--12 and 25--50 keV
energy bands for the entire set of the Aql X-1 observations are presented in
Fig.~2. These images show that 13 sources were significantly detected in the
IBIS FOV, and three of them were also covered by JEM-X.  Below, we briefly
discuss the properties of some of the observed sources.

\subsection {Aql X-1}

The Aql X-1 sky region was not accessible for observations with INTEGRAL in
the beginning of March, 2003, and therefore the rising phase of the source
outburst was not covered by INTEGRAL. The observing campaign started on
March 10, thus INTEGRAL monitored the outburst from its maximum to decay
(Fig.1). The maximum observed fluxes from Aql X-1 were $\sim 600$ and $\sim
55$ mCrab in the 5--12 keV and 25--50 keV energy bands, respectively.

Figure~3 shows two combined JEM-X$+$ISGRI spectra obtained during the first
set of observations when the Aql X-1 flux was highest and during the
fifth set of observations when the outburst was near its end. The observed
spectra can be fitted with a thermal bremsstrahlung model with temperatures 4.2
and 5.1 keV for the 1st and 5th sets, respectively. The source was
undetectable in the 50-100 keV energy band, with an upper limit of 4 mCrab
at the 68\% confidence level. However, we cannot fully rule out a presence of
a faint hard tail in the source spectrum because of the limitations of the
current data analysis discussed above. Our estimations on the flux
in tail are consist with the source spectrum of an early
stage of fast outburst decline measured by BeppoSAX (Campana et al. 1998).

\begin{figure}[t]

\vspace{-25mm}
\resizebox{115mm}{!}{\hspace{-1.2cm}\includegraphics{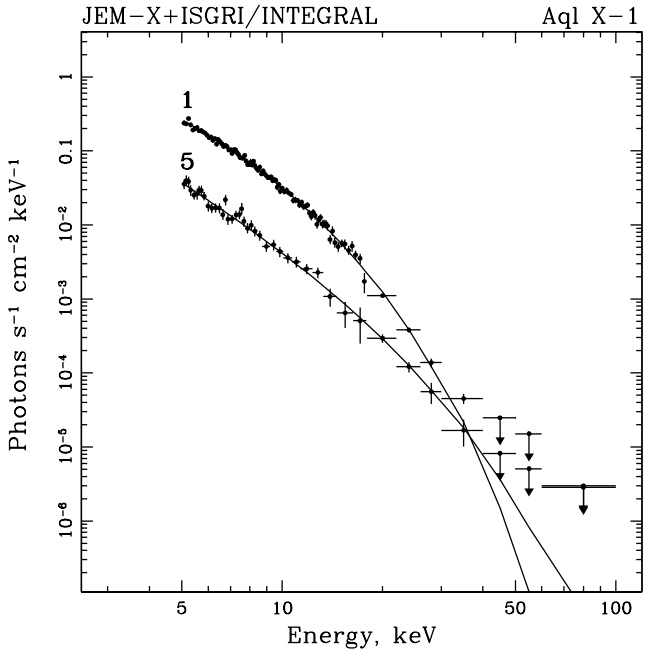}}

\vspace{-9mm}

\caption{The photon spectra of Aql X-1 observed with INTEGRAL on March 10
  (the 1st
set of observations) and on April 5 (the 5th set). The best fit thermal
bremsstrahlung models are shown by solid lines. }
\end{figure}

The JEM-X monitor has detected 2 X-ray bursts of type I from the source. The
light curve taken from the entire JEM-X detector around the moments of the bursts
is shown in the 
upper panel of Fig.4. The image reconstruction during $\sim15$~s time
intervals covering
the burst shows unambiguously (localization accuracy is $\sim4'$) that the
the bursts originated from Aql X-1 (bottom panel of Fig.4). Both the bursts have
light curves typical of the type I X-ray bursts which are considered to be
manifestations of the thermonuclear explosions on the surface of a neutron
star. Interestingly, both the bursts were detected during the 5th set of
observations when source outburst was near its end.

\begin{figure}[t]

\vspace{20mm}

\hbox{
\includegraphics[width=\columnwidth]{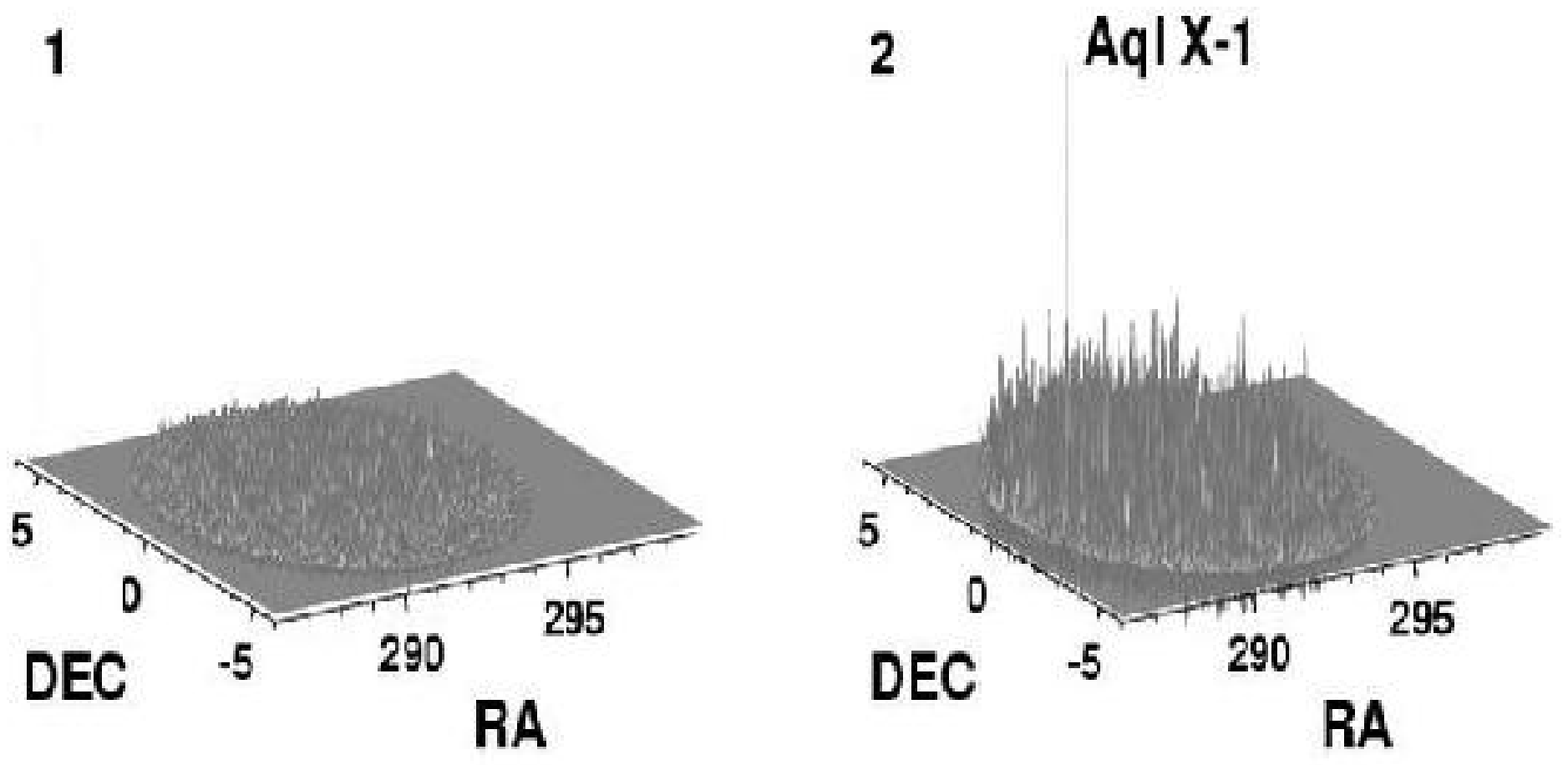}
}

\vspace{-124mm}
\resizebox{95mm}{!}{\hspace{-1.2cm}\includegraphics{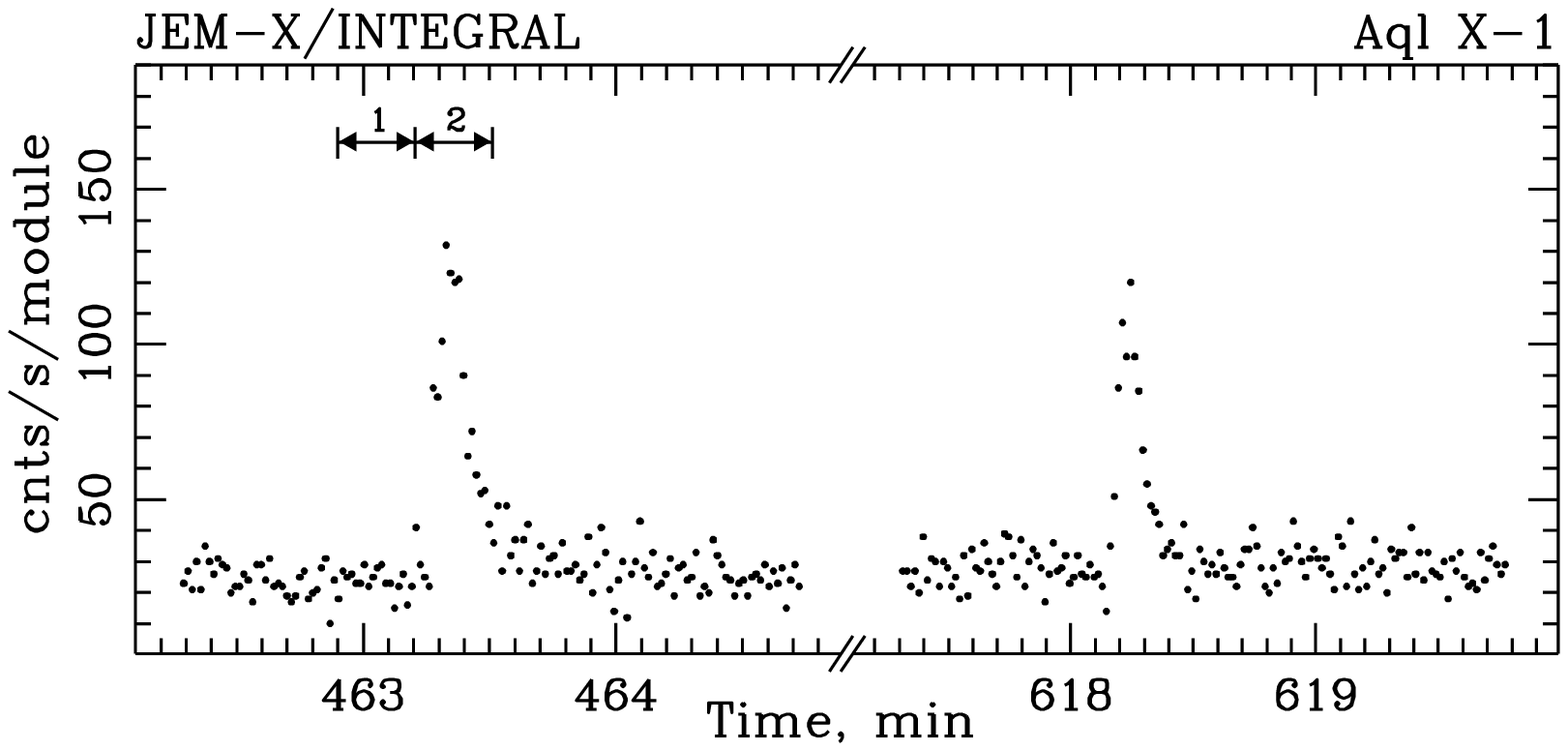}}

\vspace{24mm}

\caption{Top: JEM-X light curves containing two X-ray bursts
  from Aql X-1. Time resolution is 1 sec. The zero time on the
  horizontal axis corresponds to Apr 5.00 2003 UT. Bottom: Image
  reconstruction of the Aql X-1 field before the burst (1) and during
  the burst (2).}
\end{figure}

\subsection{X1901+03}

During all the Aql X-1 observations, the High-Mass X-ray Binary pulsar X1901+03
was detected both in the soft (JEM-X) and hard (ISGRI/IBIS) energy
bands. The preliminary source position determined with both the telescopes
agrees well with the RXTE location (Galloway et al. 2003b).

During our set of observations, the X1901+03 flux in the 25-50 keV energy
band decreased gradually from 110 mCrab to 90 mCrab. The combined JEM-X and
ISGRI/IBIS source spectrum averaged over the entire set of observations is
presented in Fig.5. The spectrum can be fited with a power law + high energy
cutoff model, with the photon index $\Gamma\sim1.95$ and the cutoff
parameters $E_{cut}\sim12$ keV, $E_{fold}\sim13.5$ keV. These values are
typical for X-ray pulsars. In addition to our observations, there is a long
set of observations of this source carried out in May, 2003, kindly made
available to us by the PIs of INTEGRAL observations of GRS 1915+105, SS 433
and Ser X-1. The source flux in May 2003 decreased by a factor of 2
relative to that in March, 2003. It would be interesting to trace the spectral
evolution of the source during the outburst decay and this analysis is now
in progress.

We performed a timing analysis of the X1901+03 JEM-X data. Fourier
transforms were computed for those periods in which X1901+03 was in the
FCFOV of JEM-X. The coherent pulsations with a frequency of $0.362$ Hz were
clearly detected along with the second harmonic. This period is consistent
with that obtained with RXTE (Galloway et al., 2003a). The period slightly
changes from the observation to observation, which may be explained by the
orbital motion of the compact object in the binary system. The light curve
obtained from the whole JEM-X detector in the 5-15 keV energy band folded
with the above pulse period is presented in Fig.6. Unfortunately, we cannot
estimate the source pulse fraction because of a lack of accurate
measurements of the JEM-X background.

\subsection{Other sources}

\begin{figure}[t]

\vspace{-25mm}
\resizebox{115mm}{!}{\hspace{-1.2cm}\includegraphics{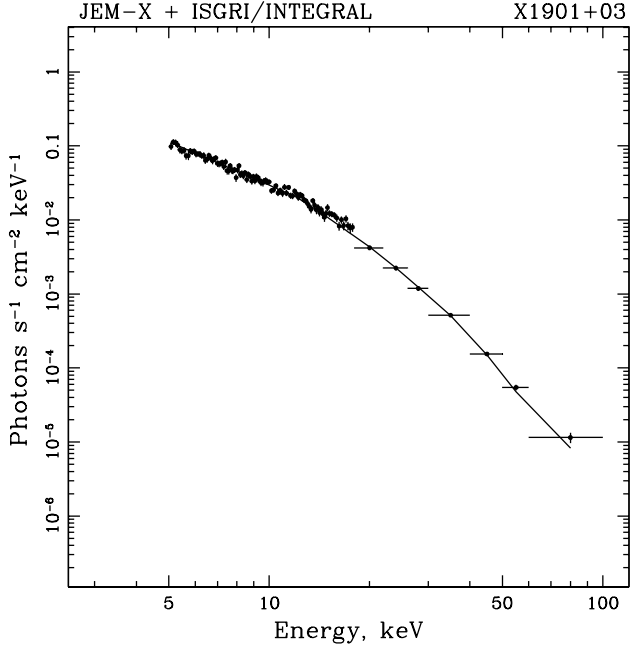}}

\vspace{-9mm}

\caption{Average photon spectrum of X1901+03 measured with JEM-X and IBIS/ISGRI
over the entire set of observations. Solid line shows the approximation
by a power law with the high energy cutoff.}
\end{figure}

\begin{figure}[t]

\vspace{-25mm}

\hbox{
\includegraphics[width=\columnwidth]{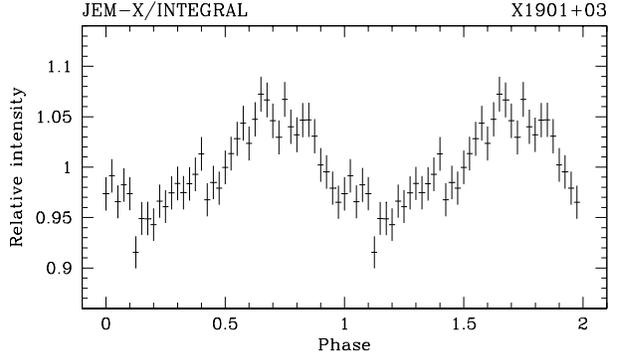}
}

\vspace{-15mm}

\caption{The pulse profile of X1901+03 in the 5-15 keV energy band. } 
\end{figure}

As was mentioned above, 13 X-ray sources were detected by the INTEGRAL
instruments during the observations of the Aql X-1 field on March-April,
2003 (Fig.2). Several of these sources (GRS1915+105, SS443, Ser X-1, and
IGR J19140+098) are subjects of other INTEGRAL proposals and therefore are
not discussed here.  We present the averaged hard (20--50~keV) X-ray fluxes
for the remaining sources in IBIS FOV (Table~\ref{tab:fluxes}). 

\begin{table}[h]
  \centering
  \caption{Hard X-ray fluxes for sources in the Aql X-1 field}
  \label{tab:fluxes}
  \begin{tabular}{lr}
    XTE J1855-026 & $\sim13$ mCrab \\
    4U1907+097 & $\sim12$ mCrab \\
    X1908+075  & $\sim17$ mCrab \\
    XTE J1908+094  & $\sim9.5$ mCrab \\
    X1916-053  & $\sim8$ mCrab \\
    IGR J18483-0311 & $\sim13$ mCrab \\
  \end{tabular}
\end{table}

Several
points are noteworthy: 1) sources fluxes are variable by a factor of a few,
but here we quote only the average values; 2) the fluxes were derived
assuming that the Crab intensity in the 25--50 keV energy band of ISGRI
detector is 130 counts/s; 3) the new transient source IGR J18483-0311 was formally
discovered with INTEGRAL on Apr 23-28, 2003 (Chernyakova et al. 2003), but
it is also detectable in one of our observations (Apr 5, 2003) with a signal
to noise ratio of $S/N\sim10$ in the 25-50 keV energy band.

A detailed analysis of the sources in the Aql X-1 field is beyond the scope
of this paper. Here we discuss only the preliminary spectral analysis for
three sources of different classes -- 4U1907+097, XTE J1908+094, and
X1908+075. The first of them is a typical X-ray pulsar with the energy
spectrum described by a power law with a high energy cutoff. The cyclotron
absorption line with several harmonics was detected in the spectrum of this
source (Coburn et al. 2002).  The second source is a likely black hole
candidate with a relatively hard, power law spectrum ($\Gamma\sim 2$)
without any indication for a high energy cutoff. No pulsations or type I
X-ray bursts were observed from XTE J1908+094 (in't Zand et al. 2002). The
nature of the third source, X1908+075, is not yet determined.

The average ISGRI photon spectra of 4U1907+097, XTE J1908+094
and X1908+075 in the 18-100 keV energy band are presented in Fig.7.  In all
three cases, the spectra can be described well by a power law model with
the photon index $\Gamma\sim 2.1$ for XTE J1908+094, $\sim 2.9$ for
X1908+075, and $\sim 3.9$ for 4U1907+097. The spectral slope of XTE
J1908+094 is similar to that of other accreting BHs in the 
low/hard state and the value of its photon index $\Gamma$ is in agreement
with the previous results of the BeppoSAX observatory (in't Zand et
al. 2002). The photon index for 4U1907+097 is typical of X-ray pulsars
spectra in the hard energy band. We note that this spectrum can be equally
well described by the power law + high energy cutoff model. The power law
and cutoff parameters cannot be derived independently from the INTEGRAL data
due to lack of data below 18 keV. Therefore, we fixed some of the parameters
(the photon index $\Gamma$ and cutoff energy $E_{cut}$) at the values
obtained by the RXTE observatory and then derived $E_{fold}\simeq10$ keV, in
agreement with the RXTE results $E_{fold}=9.8\pm0.6$ keV (Coburn et
al. 2002).

The spectrum of X1908+075 has the intermediate value of the photon index
with respect to typical black hole binaries (e.g., XTE J1908+094) and the
X-ray pulsars (4U1907+097). No obvious cyclotron features are present in
the source spectrum, but they cannot be strongly ruled out because of
the limited spectral resolution of the ISGRI instrument. Therefore, the
question about the nature of X1908+075 is still open.

\begin{figure}[t]
\includegraphics[width=\columnwidth,bb=90 260 465 700,clip]{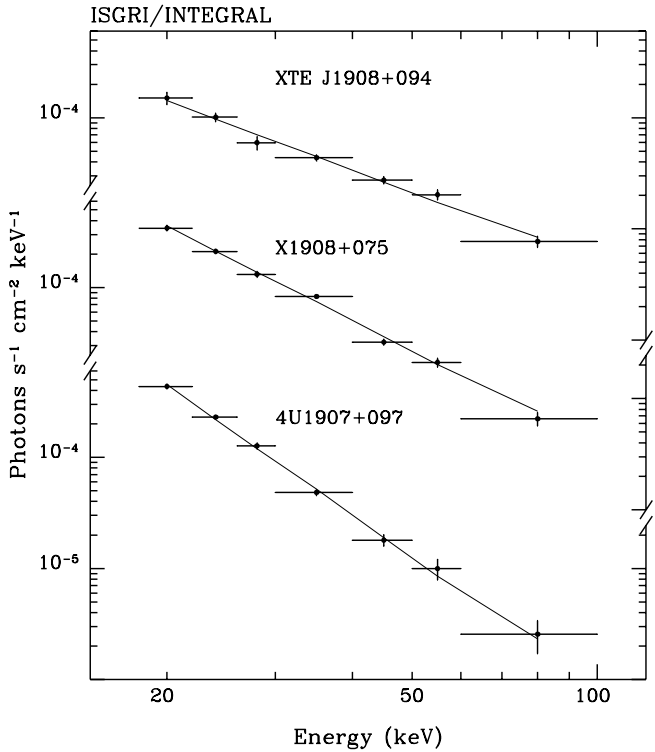}

\caption{Average photon spectra of XTE J1908+094, X1908+075 and 4U1907+097
obtained with the ISGRI instrument. Solid lines show the power law
approximations. 
\label{spec_other}}
\end{figure}

\section{Conclusions}

The transient X-ray source Aql X-1 was observed with INTEGRAL during the
source outburst in March-April 2003. The observing strategy allowed us to
cover almost the entire outburst episode, and therefore investigate the
source in the states with different luminosity. The preliminary spectral
analysis indicates that the spectrum in the 3-40 keV energy band can be
described by a thermal bremsstrahlung model. The source was undetectable
in the harder energy band. We observed two X-ray bursts of type I from Aql
X-1.

Dozen other sources were detected in the INTEGRAL field of
view. We reconstructed the pulse profile and average spectrum of one of the
most interesting of these sources, the pulsar X1901+03 (indentified as a
pulsar only one month before our observations, see above). The 
18-100 keV energy spectra of three other bright sources, XTE J1908+094,
X1908+075, and 4U1907+097 can be described by a simple power law model with
photon indices of $2-3.9$.

A more detailed analysis of the sources in the Aql X-1 field is in progress.

\bigskip

{\it Acknowledgements.} We would like to thank Mike Revnivtsev and
Alexey Vikhlinin for very
useful discussions and comments.  
This research has made by using of data obtained through the INTEGRAL
Science Data Center (ISDC). 
This work was supported by RFBR grant 03-02-06772, grants of
Minpromnauka NSH-2083.2003.2 and 40.022.1.1.1102, and the program
of the Russian Academy of Sciences ``Nonstationar phenomena in astronomy''.

\end{document}